\documentclass[11pt, oneside,fleqn]{article}   	
\usepackage{geometry}                		
\usepackage{graphicx}				
\usepackage{amssymb}

 \usepackage{helvet}

\usepackage{wrapfig}

\usepackage{float}

\title{SIR Model with Delay Yielding a Logistic Solution}
\author{Paul A. Reiser (reiser.paul@gmail.com)}

\begin{document}
\maketitle

\section{Abstract}


The classical SIR pandemic model suffers from an unrealistic assumption: The rate of removal from the infectious class of individuals is assumed to be proportional to the number of infectious individuals. This means that a change in the rate of infection is simultaneous with an equal change in the rate of removal. A more realistic assumption is that an individual is removed at a certain time interval after having been infected. A simple modified SIR model is proposed which implements this delay, resulting in a single delay differential equation which comprises the model. A solution to this DDE which is applicable to a pandemic is of the form $A+B L(t)$ where $L(t)$ is a logistic function, and $A$ and $B$ are constants. While the classical SIR model is perhaps an oversimplification of pandemic behavior, it is instructive in that many of the fundamental dynamics and  descriptors of pandemics are clearly and simply defined. The logistic model is generally used descriptively, dealing as it does with only the susceptible and infected classes and the rate of transfer between them. The present model presents a full but modified SIR model with a simple logistic solution which is more realistic and equally instructive.

\section{Introduction}

The classical SIR model consists of three classes or compartments\cite{K&M1927,Hethcote2000}:

S(t) - The fraction of the population which has never been infected, and is therefore susceptible to infection.

I(t) - The fraction of the population which is "infectious" - members of this compartment have been infected, but have not yet become immune (or died), and are capable of passing the infection on to any susceptible person that they come in contact with.

R(t) - The fraction of the population which has been removed from the susceptible and infectious compartments. These are individuals who have either become immune or have died from the disease. Removal is irreversible.

In the following discussion, subscripted variables of the form $t_x$ will represent a particular time, $T_x$ will represent a particular time interval and $f_x$ will represent the associated frequency: (i.e. $T_x f_x=1$). It is assumed that birth and  death rates have an insignificant effect on the total population, which will be assumed constant so that $S(t)+I(t)+R(t)=1$. There are two assumptions made in the classical SIR model concerning the rates of transfer between compartments. The first assumption is that the fractional rate of transfer of individuals from the susceptible class to the infected class ($v(t)$) is given by:

$v(t)=f_c \,S(t) I(t)$

\noindent  where  $f_c$  is the average frequency of potentially infectious contacts experienced by an individual member of the population. A potentially infectious contact is one which, if the contacting pair consists of a susceptible and an infectious individual, the susceptible individual will certainly become infected.  

The second assumption is that the probability of an infectious individual moving from the infectious class to the removed class in an infinitesimal time interval $dt$ is $f_r\,dt$ where $f_r$ is a constant. This probability is thus independent of the time since infection. The rate of removal is then proportional to the fraction of infectious individuals divided by the average infectious period $T_r=1/f_r$. The SIR model is then:

 $S'(t)=-f_c \, S(t) I(t) $
 
 $I'(t)=f_c  \,S(t) I(t) - f_r \,I(t) $
 
 $R'(t)=f_r \, I(t) $

Solutions to this model are complicated by the non-linearity of the above differential equations.\cite{K&M1927,Harko2014,Prodanov2021} 

\section{SIR model with delay}

A problem with the classical SIR model is that, via the $f_r I(t)$ term in the expression for $I'(t)$, a spike in the infection rate results in an immediate spike in the recovery rate, which is unrealistic.\footnote{Compartmental models with time delay have been implemented previously  \cite{Ma2004}-\cite{Wang2002b} in order to obtain more realistic results, but without implementing a delay in the transition from the infectious to the recovered compartment.} This results from the unrealistic assumption that the probability of removal during a particular time interval is independent of the time since infection.  In the present model,  it is assumed that an individual spends a fixed amount of time $T_r$ in the infectious compartment. A spike in the infection rate will result in a equal spike in the recovery rate at time $T_r$ later.  In other words, the number of recovered individuals at time $t$ is equal to the number of recovered individuals at time $t-T_r$ plus the number of infected individuals at time $T_r$ because those infected individuals will be the only individuals contributing to the set of recovered individuals during the time interval between $t-t_r$ and $t$. The rate of removal at time $t$ is therefore equal to the rate of infection at time $t-T_r$.




To model this situation, it is useful to define $n(t)$, the cumulative fraction of individuals who have ever been infected ($n(t)=I(t)+R(t)$). The fractional populations may then be written as:

$S(t)=1-n(t)$

$I(t)=n(t)-n(t-T_r)$

$R(t)=n(t-T_r)$

\noindent The rate of infection is again assumed to be:

$v = n'(t) = f_c \,S(t)\, I(t)$

\noindent The model is then described by a single delay differential equation:

\begin{equation}\label{main}
n'(t)= f_c  \,S(t)  \,I(t) =  f_c \,[1-n(t)]  \, [n(t)-n(t-T_r)]
\end{equation}
\noindent the solution to which will allow the S, I, and R functions to be found. This equation involves two "pandemic parameters" which describe the dynamics of the pandemic, $f_c$ and $T_r$.  A pandemic model will usually assume as an initial condition that $n(-\infty)=n_m$ which is constant and often set to zero unless studying an outbreak in a partially immune population. The model will reach a steady state solution  $n(\infty)=n_p$. There will also be a constant of integration $t_h$ which specifies the "center" of the pandemic.  As shown below, Equation \ref{main} then has the following logistic solution:
\begin{equation}\label{mainsoln}
n(t) =n_m+\frac{n_p-n_m}{1+e^{-f_e(t-t_h)}}
\end{equation}
\noindent where the phenomenological parameters\footnote{ Phenomenological parameters may be obtained, for example, by fitting the logistic solution to measured pandemic data.} $n_p$ and $f_e$ are functions of the pandemic parameters $f_c$ and $T_r$ and initial condition $n_m$. Defining the basic reproduction number\cite{Hethcote2000}:

$R_o = f_c T_r (1-n_m)$

\noindent the phenomenological parameters in Equation \ref{mainsoln} are given by:
\begin{equation}\label{Te}
f_e =f_r(R_o+W_0(-R_o e^{-R_o}))
\end{equation}
\begin{equation}\label{np}
n_p=n_m+f_e/f_c
\end{equation}
\noindent where $W_j(z)$ is the $j$-th branch of the Lambert $W$ function\cite{Corliss1996}, also known as the "product log" function. The branch parameter $j$ has been set to zero, since only for $j=0$ will $f_e$ be real and finite. From the properties of the $W$ function, Equation \ref{Te} will yield $f_e=0$ for $R_o \le 1$, implying that $n(t)=n_m$ and that the infectious agent is unable to propagate.  $n_p=n(\infty)$ is the fraction that have ever been infected at infinite time (disease-free equilibrium). $t_h$ is the "half way point": The time at which $n(t)=(n_p+n_m)/2$ and it is an initial condition. $t_h$ is also the inflection point of $n(t)$ and so specifies the peak rate of infection. $f_e$ is the time scaling parameter, with $T_e=1/f_e$ being a measure of the duration of the pandemic.

Note that the standard SIR model can be recovered from this model by assuming that $T_r$ is so small that the following approximation may be made:

$n'(t-T_r)\approx \frac{n(t)-n(t-T_r)}{T_r} = f_r\, I(t)$

\section{Expression for $f_e$ in terms of $f_c$ and $T_r$}

The scaling parameter $f_e$ can be found by considering the situation near time plus or minus infinity. As $t\rightarrow -\infty$, $e^{f_e(t-t_h)}$ becomes small, and the main equation \ref{main} to first order is:

$ n'(t)\rightarrow f_c \,[1-n_m]\,[n(t)-n(t-T_r)]$ 

\noindent which can be solved:

$ n(t)\rightarrow n_m +(n_p-n_m)\, e^{f_e(t-t_h)}$ 

\noindent where $f_e$ obeys:
\begin{equation}\label{fTe1}
f_e = f_c(1-n_m)(1-e^{-f_e T_r})
\end{equation}
\noindent the solution to which yields Equation \ref{Te}. 

\section{Expression for $n_p$ in terms of $f_c$ and $T_r$}

A similar analysis at near infinite time yields:

$ n(t)\rightarrow n_p - (n_p-n_m)\, e^{-f_a(t-t_h)}$ 

\noindent From the symmetry of the logistic function, it can be seen that $f_a=f_e$ so that:
\begin{equation}\label{fTe2}
f_e = f_c(1-n_p)(e^{f_e T_r}-1)
\end{equation}
\noindent Assuming $f_e$ is positive and finite, the above equation has solution:

$f_e T_r = -R_p - W(-1,-R_p\, e^{-R_p})$

\noindent where $R_p= f_c T_r (1-n_p)$. The $j=-1$ branch of the $W$ function has been chosen since it is the only one for which $f_e$ will be real and finite.  Equation \ref{np}  now follows from Equations \ref{fTe1} and \ref{fTe2}. Note that the pandemic parameters $f_c$ and $T_r$ may in turn be expressed in terms of  phenomenological parameters $f_e$ and $n_p$ and initial condition $n_m$ by:
\begin{equation}\label{Tc}
f_c=\frac{f_e}{n_p-n_m}
\end{equation}
\begin{equation}\label{Tr}
T_r=\frac{1}{f_e}\ln\left(\frac{1-n_m}{1-n_p}\right)
\end{equation}

\section{Proof of solution}
Supressing the time dependencies and defining:

$\epsilon=e^{-f_e(t-t_h)}$

\noindent Eq. \ref{mainsoln} may be solved for $\epsilon$:

\begin{equation}\label{E1}
\epsilon=\frac{n_p-n}{n-n_m}
\end{equation}

\noindent and the derivative of Eq. \ref{mainsoln} may be expressed as:

$
n' =\frac{f_e}{n_p-n_m} (n-n_m)(n_p-n)
$

\noindent The logistic solution will be a solution to the pandemic model if it can be shown that the above derivative is equal to the derivative given in the model Equations \ref{main}. Using Equation \ref{Tc} and defining $n_r=n(t-T_r)$, it will be required that:
\begin{equation}\label{E0}
(n-n_m)\,(n_p-n) = (1-n)\,(n-n_r)
\end{equation}

\noindent Define $\epsilon_r=e^{f_e T_r}$. Using Equations \ref{fTe1} and \ref{np} it follows that:

\begin{equation}\label{E2}
\epsilon_r=\frac{1-n_m}{1-n_p}
\end{equation}

\noindent From Eq. \ref{mainsoln}, it will then be required that:
  
\begin{equation}\label{E3}
n(t-T_r) = n_r = n_m+\frac{n_p-n_m}{1+\epsilon_r \epsilon}
\end{equation}

\noindent With $\epsilon$, $\epsilon_r$, and $n_r$ now expressed in terms of $n$, $n_p$ and $n_m$ by equations \ref{E1}, \ref{E2} and \ref{E3}, simple substitution and algebraic manipulation will demonstrate that Equation \ref{E0} and therefore Equation \ref{E3} are in fact true.


\end{document}